\documentclass[preprints,article,accept,moreauthors,pdftex]{Definitions/mdpi} 

\firstpage{1} 
\pubvolume{xx}
\issuenum{1}
\articlenumber{5}
\pubyear{2019}
\copyrightyear{2019}
\history{}

\usepackage{adjustbox}
\usepackage{txfonts}

\Title{Detectable Optical Signatures of QED Vacuum Nonlinearities using High-Intensity Laser Fields}

\newcommand{\newT}[1]{#1}
\newcommand{\newTT}[1]{{#1}}

\Author{Leonhard Klar}

\AuthorNames{Firstname Lastname, Firstname Lastname and Firstname Lastname}

\address{%
Theoretisch-Physikalisches Institut, Abbe Center of Photonics, Friedrich-Schiller-Universit\"at Jena, Max-Wien-Platz 1, 07743 Jena, Germany;\\
Helmholtz-Institut Jena, Fr\"obelstieg 3, 07743 Jena, Germany; leonhard.klar@uni-jena.de\\
}

\abstract{Up to date, quantum electrodynamics (QED) is the most precisely tested quantum field theory. Nevertheless, particularly in the high-intensity regime it predicts various phenomena that so far have not directly been accessible in all-optical experiments, such as photon-photon scattering phenomena induced by quantum vacuum fluctuations. Here, we focus on all-optical signatures of quantum vacuum effects accessible in the high-intensity regime of electromagnetic fields. We present an experimental setup \newT{giving rise to} signal photons distinguishable from the background. This configuration \newT{ is based} on two optical pulsed petawatt lasers\newT{: one} generates a narrow but high-intensity scattering center to be probed by the other one. We calculate the differential number of signal photons attainable \newT{with} this field configuration analytically and compare it with the background of the driving laser beams.
}

\begin{document}

\section{Introduction}

Shortly after Dirac predicted the positron and introduced his idea of the Dirac-Sea \cite{Dirac28a,Dirac28b,Dirac30}, Sauter used his theory to describe the creation of an electron-positron pair in presence of a strong electromagnetic field \cite{Sauter31}. In the 30s of the 20th century, Heisenberg and Euler formulated a Lagrangian -- the famous Heisenberg-Euler-Lagrangian $\mathcal{L}_{\mathrm{HE}}$ -- that averages over the virtual electron-positron fluctuations. \newT{The latter predicts} nonlinear self-interaction \newT{of} electromagnetic fields in \newT{the quantum vacuum, facilitating} photon-photon-scattering \newT{phenomena} \cite{Euler35,Heisenberg36,Karplus51}.

A relevant scale in the Heisenberg-Euler Lagrangian is the critical field strength $\mathcal{E}_{\mathrm{crit}}=c^3m_e^2/\left(e\,\hbar\right)\approx1.3\times10^{18}\,\mathrm{V\,m^{-1}}$ \newT{or} $B_{\mathrm{crit}}=\mathcal{E}_{\mathrm{crit}}/c\approx 4\times10^9\,\mathrm{T}$\newT{, respectively}. Here $m_e$ is the electron mass, $e$ the elementary charge, $c$ the speed of light, and $\hbar$ Planck's reduced constant. We characterize a field as strong if it approaches the order of magnitude of this threshold. \newT{Due to the large advances in } laser technology during the last decades, \newT{it might become possible to find } signatures of quantum vacuum nonlinearities in experiments \newT{with strong laser fields in the near future. Various} phenomenona of \newT{quantum vacuum} nonlinearity, e.g. photon-photon scattering, vacuum birefringence, quantum reflection, photon splitting, and more, appear to be detectable with state-of-the-art lasers \cite{BialynickaBirvla70,Reinhardt77,Moulin99,Dittrich00,Dunne04,Heinzl06,Lundstrom06,Marklund08,Heinzl09,Piazza12,Battesti12,Gies13,Gies15,Karbstein15a,King16a,King16b,Karbstein16b,Inada17,Karbstein17,Meuren17,Blinne18a,Gies18a,Gies18b,Kohlfuerst18,Bulanov19}.

In this work, we focus on photon-photon scattering as a signal of \newT{effective} nonlinear interactions \newT{of electromagnetic fields} mediated by quantum fluctuations. We use the Heisenberg-Euler-Lagrangian $\mathcal{L}_{\mathrm{HE}}$ to obtain an analytic \newT{expression} for the density of signal photons by utilizing the emission picture \newT{at} one-loop order. Furthermore, \newT{to simplify our calculations} we restrict ourselves to Gaussian beams in the limit of infinite Rayleigh lengths. \newT{As a means to enhance the signal} we suggest a laser setup with two high-intensity lasers, one of which \newT{is} split into three different pump beams \newT{of} different frequencies. In section \ref{sec:geo} we explain this configuration and study the \newT{attainable signals} in the following section \ref{sec:results}. We \newT{derive} the differential number of signal photons and compare these results with the background \newT{constituted by} the driving laser beams. Ultimately, we \newT{show how to} generate a spatially localized scattering center which leads to signal photons scattered wide enough to be distinguishable from the background photons. 
 
\section{Theoretical Background} \label{sec:Theory}
In the following steps we use the Heaviside-Lorentz system with natural units $\left(\hbar=c=1\right)$. Our metric convention is $g_{\mu\nu}=\mathrm{diag}\left(-,+,+,+\right)$. 

For describing \newT{the} QED \newT{vacuum} including vacuum fluctuations we use the Heisenberg-Euler-Lagrangian, $\mathcal{L}_{\mathrm{HE}} = \mathcal{L}_{\mathrm{MW}} + \mathcal{L}_{\mathrm{NL}}$, where $\mathcal{L}_{\mathrm{MW}}=-\left(1/4\right) F^{\mu\nu}F_{\mu\nu}$ \newT{denotes the Maxwell Lagrangian} with the field strength tensor $F^{\mu\nu}$ \newT{and} $\mathcal{L}_{\mathrm{NL}}$ \newT{accounts for higher-order, non-linear terms in $F^{\mu\nu}$} extending Maxwell\newT{'s linear theory in vacuum} \cite{Schwinger51,Euler35,Heisenberg36}. We want to focus on signal photons created by \newT{these} nonlinearities of the QED vacuum. To describe them we choose the vacuum emission picture \cite{Karbstein15b,Karbstein15c,Karbstein15d,Gies18b,Karbstein19b}. In order to obtain \newT{a sizable amount of} these photons we need a strong background field which we denote with $\bar{F}^{\mu\nu}$; additionally, the absolute value of this field is denoted by $\bar{F}$. \newT{The Heisenberg-Euler-Lagrangian depends on} the background fields \newT{via} the invariant quantities 
\begin{equation}
\mathcal{F}=\frac{1}{4} \bar{F}^{\mu\nu}\bar{F}_{\mu\nu}= \frac{1}{2}\left(\mathbf{B}^2-\boldsymbol{\mathcal{E}}^2\right) \qquad \text{and} \qquad \mathcal{G}=\frac{1}{4} \tilde{\bar{F}}^{\mu\nu}\bar{F}_{\mu\nu}= -\mathbf{B}\cdot\boldsymbol{\mathcal{E}}\,, \label{eq:invariantFG}
\end{equation}      
using the dual field strength tensor $\tilde{F}^{\mu\nu}=-1/2\,\varepsilon^{\mu\nu\alpha\beta}F_{\alpha\beta}$ and the vector representation of the electric field strength $\boldsymbol{\mathcal{E}}$ and magnetic field strength $\mathbf{B}$. We use the one-loop and lowest-order expansion of the nonlinear term of the Heisenberg-Euler effective Lagrangian,
\begin{equation}
\mathcal{L}_{\text{eff}} = \frac{2}{45} \frac{\alpha^2}{m_e^2} \left(4 \mathcal{F}^2 + 7 \mathcal{G}^2 \right) + m_e^4\; \mathcal{O} \label{eq:Leff}\left(\left(\frac{\alpha \bar{F}^2}{m_e^4}\right)^3\right)\,,
\end{equation}
with the fine-structure-constant $\alpha=e^2/\left(4\pi\right)\approx1/137$ \cite{Euler35,Heisenberg36,Karbstein16b}. Obviously, the corresponding diagrams are
\begin{equation}
\mathcal{L}_{\text{eff}} = \adjustimage{valign=c}{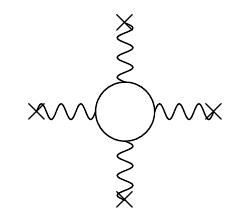} + \adjustimage{valign=c}{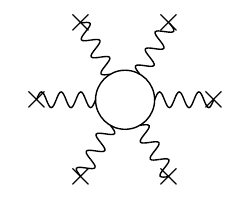} +\quad \dots \label{eq:Leff_feyn}
\end{equation}
and contain only even numbers of external photons, according to Furry's theorem \cite{Furry37a}. The leading order in Eq. \eqref{eq:Leff_feyn} is the coupling of four photons \newT{via a virtual electron-positron vacuum fluctuation}; all higher orders will be suppressed by powers of $\alpha\bar{F}^2/m_e^4 \propto \bar{F}^2/\mathcal{E}_{\text{cr}}^2$.

In order to count the number of signal photons in the vacuum emission picture for \newT{the} setup described in section \ref{sec:geo}, it is necessary to evaluate the signal photon amplitude $S_{\left(p\right)}\left(\mathbf{k}\right)$. This is the scattering amplitude \newT{from the vacuum state to} one \newT{signal} photon $\gamma_{\left(p\right)}\left(\mathbf{k}\right)$ with polarization $\left(p\right)$ and three dimensional wave vector $\mathbf{k}=\mathrm{k}\,\hat{\mathbf{k}}$ with $\hat{\mathbf{k}}= \left(\cos{\varphi}\,\sin{\vartheta},\, \sin{\varphi}\,\sin{\vartheta},\,\cos{\vartheta}\right)$. We can determine the signal photon amplitude as \cite{Gies18b}
\begin{align}
S_{\left(p\right)}\left(\mathbf{k}\right) &= \left<\left.\left.\gamma_{\left(p\right)}\left(\mathbf{k}\right)\right|\Gamma_{\text{int}}\left[\bar{A}\left(x\right)\right]\right|0\right> \nonumber \\
&\newTT{\stackrel{\text{LCFA}}{\approx}    \mathrm{i}  \frac{\epsilon^{*\mu}_{\left(p\right)}\left(k\right)}{\sqrt{2k^0}} \int \mathrm{d}^4x\; \mathrm{e}^{\mathrm{i}k_{\alpha}x^{\alpha}} \left( k^{\nu} \bar{F}_{\nu\mu} \frac{\partial \mathcal{L}_{\text{eff}}}{\partial \mathcal{F}}  + k^{\nu} \tilde{\bar{F}}_{\nu\mu} \frac{\partial \mathcal{L}_{\text{eff}}}{\partial \mathcal{G}} \right) }\,, \label{eq:S_p_Gamma}
\end{align} 
where $\Gamma_{\text{int}}\left[\bar{A}\left(x\right)\right]$ is the \newT{effective} action \newT{governing} the nonlinear interaction \newT{of electromagnetic fields characterized by the} electromagnetic vector potential $\bar{A}\left(x\right)$ \newT{and $\epsilon^{*\mu}$ denotes the polarization of the signal photons with wave vector $\mathbf{k}$. Note that $\mathrm{k}=k^0=\sqrt{k_x^2+k_y^2+k_z^2}$}. \newT{The typical spatial and temporal scales characterizing} the driving laser beams \newT{are}  much \newT{larger} than the reduced Compton wavelength $\lambdabar_{\mathrm{C}}=1/m_e\approx 3.86\times10^{-13}\,\mathrm{m}$ \newT{and} Compton time $\tau_{\mathrm{C}}\approx1.29\times 10^{-21}\,\mathrm{s}$ \newT{of the electron, respectively}. This justifies to use the locally constant field approximation (LCFA) \cite{Karbstein15a,Karbstein15b,Karbstein16b,Gies18b} \newT{ adopted in the second line of Eq. \eqref{eq:S_p_Gamma} }.

In the LCFA, \newT{$S_{\left(p\right)}\left(\mathbf{k}\right)$} is determined by the derivatives of the effective one-loop Heisenberg-Euler Lagrangian 
\begin{equation}
\left\{\begin{array}{c}
\frac{\partial \mathcal{L}_{\text{eff}}}{\partial \mathcal{F}} \\
\frac{\partial \mathcal{L}_{\text{eff}}}{\partial \mathcal{G}}
\end{array}\right\} = \frac{e^2}{4\pi} \frac{1}{45} \left(\frac{e}{m_e^2}\right)^2 \left\{\begin{array}{c}
4 \mathcal{F}\left(x\right)\\
7 \mathcal{G}\left(x\right)
\end{array}\right\} + \mathcal{O}\left(\left(\frac{e \bar{F}}{m_e^2}\right)^4\right)\,,
\end{equation}
where the fine-structure-constant $\alpha$ is expressed via the elementary charge $e$. In the limit of weak electromagnetic fields -- weak compared to the critical field strength $\mathcal{E}_{\text{crit}}$ -- we neglect higher-order terms of $\mathcal{O}\left(\left(e \bar{F}/m_e^2\right)^4\right)$, and the signal photon amplitude \newT{can be expressed as} 
\begin{equation}
S_{\left(1\right)} \left(\mathbf{k}\right) = \frac{1}{\mathrm{i}}  \frac{e^2}{4\pi}  \frac{m_e^2}{45} \sqrt{\frac{k^{0}}{2}} \left(\frac{e}{m_e^2}\right)^3 \int \mathrm{d}^4x\; \mathrm{e}^{\mathrm{i}k_{\alpha}x^{\alpha}} \left( 4\left[\mathbf{e}_{\left(1\right)}\cdot\boldsymbol{\mathcal{E}} - \mathbf{e}_{\left(2\right)}\cdot \mathbf{B}\right] \mathcal{F}  + 7 \left[\mathbf{e}_{\left(1\right)}\cdot\mathbf{B} + \mathbf{e}_{\left(2\right)}\cdot \boldsymbol{\mathcal{E}}\right] \mathcal{G} \right)\,, \label{eq:S1_FG}
\end{equation}
and $S_{\left(2\right)}\left(\mathbf{k}\right) = \left. S_{\left(1\right)}\left(\mathbf{k}\right)\right|_{\substack{\mathbf{e}_{\left(1\right)}\rightarrow\mathbf{e}_{\left(2\right)}\\\mathbf{e}_{\left(2\right)}\rightarrow-\mathbf{e}_{\left(1\right)}}}$. Here we have introduced the unit vectors $\mathbf{e}_{\left(p\right)}$ \newT{with $p\in\left\{1,2\right\}$}, which \newT{span the polarizations} of the signal photon. We define them by $\mathbf{e}_{\left(1\right)}=\left(\cos\varphi\,\sin\vartheta,\,\sin\varphi\,\cos\vartheta,\,-\sin\vartheta\right)$ and $\mathbf{e}_{\left(2\right)}=\left(-\sin\varphi,\,\cos\varphi,0\right)$.

\section{Geometrical setup} \label{sec:geo}
We suggest a special collision geometry \newT{of the driving laser pulses} generating a tightly focused field configuration. \newT{For later references, we distinguish between pump and probe laser fields. The superposition of several pump pulses results in a narrow strongly peaked field region with is probed by the counter} \newTT{propagating} \newT{ probe beam.} Here we consider two high-intensity optical laser beams, each with a photon energy $\omega_0=2\pi/\lambda=1.55\,\mathrm{eV}$. In SI units the associated wavelength is $\lambda=800\,\mathrm{nm}$. Both lasers belong to the petawatt class and deliver a pulse duration of $\tau=25\,\mathrm{fs}$, focused to a beam waist size $w_{i}=\lambda$. \newT{For the probe laser we assume a total pulse energy of $W=25\,\mathrm{J}$ and for the pump pulse a total energy of $W_{\text{pump}}=50\,\mathrm{J}$. As noted above, the latter will be partitioned into several pulses.}  Laser facilities \newT{providing beams of such energies} are available by now \cite{Gies18b,Walker99,Matras13,Rus17}. The \newT{peak} field strength $\mathcal{E}_{\star}$ \newT{associated with a pulse energy $W=25\,\mathrm{J}$ is}
\begin{equation}
\mathcal{E}_{\star} = \sqrt{2\sqrt{\frac{2}{\pi}} \frac{W \omega_0^2}{\pi^3 \tau}} \approx 1.1 \times 10^{15}\frac{\mathrm{V}}{\mathrm{m}}\,, \label{eq:E_star}
\end{equation}
and satisfies the approximations done in section \ref{sec:Theory}.

\newT{The} pulsed laser \newT{of pulse energy \newT{$W_{\text{pump}}=50\,\mathrm{J}$} constitutes the} pump field. Instead of limiting \newT{ourselves} to \newT{a single pump} beam we use it to generate a high-intensity localized field configuration \newT{by splitting it into three parts which are subsequently superimposed, thereby producing a particularly strong field in the common beam focus.} This composition can be \newT{achieved} by using optical mirrors or beam splitters before focusing \cite{Thaury07}.
Furthermore, we want to equip all three colliding pump beams with different frequencies\newT{, i.e., we want to achieve $\omega_0 \rightarrow \nu_i \omega_0$, where $\nu_i$ denotes a natural number; see below.} Experimentally, high-harmonic generation is one way to realize \newT{several beams of } different frequencies \newT{from a single driving beam}. This leads us to introduce frequency factors $\nu_i$ which are $\nu_1=1$, $\nu_2=2$ and $\nu_3=4$. We focus on three pump lasers plus one additional probe laser; therefore we label the probe laser with $i=0$ and the pump laser with $i\in\left\{1,2,3\right\}$. \newT{Each higher-harmonic generation implies losses; for the frequency doubling process conserving the pulse duration $\tau$, the loss factor can be estimated as 59.55\%, as evidenced experimentally in \cite{Marcinkevicius04}.} \newT{Hence, when aiming at} using this technique to generate a strong confined electromagnetic field it is indispensable to account for \newT{losses of the pulse energy in the conversion process.}
\newT{In line with the above estimate of the loss factor, we assume a conversion efficiency of the pulse energy of 40.45\%} for every high-harmonic generation including mirrors and splitters. 
\newT{The first pump laser keeps its frequency and hence pulse energy resulting in an effective pulse energy of $W_1^{\text{eff}}=25\,\mathrm{J}$. We divide the remaining pump pulse energy into two pules with $W_2=15.55\,\mathrm{J}$ and $W_3=9.55\,\mathrm{J}$. Note, however, after frequency doubling only an effective pulse energy of $W_2^{\text{eff}}=6.25\,\mathrm{J}$ remains for the second pump laser and $W_3^{\text{eff}}=1.5625\,\mathrm{J}$ for the third, respectively. We can convert theses different pulse energies to the corresponding field strength amplitudes, see Eq. \eqref{eq:E_star}, and determine relative amplitudes $A_i$ measuring theses fields in terms of the peak field strength $\mathcal{E}_{\star}$. This results in $A_1=1$, $A_2=0.5$, and $A_3=0.25$. We use theses amplitudes in the subsequent section to introduce a general expression for the field profile $\mathcal{E}_i\left(x\right)$; see Eq. \eqref{eq:E_x_infRay}.} The probe laser \newT{is left} unaltered, implying $W=25\,\mathrm{J}$, $A_0=1$ and $\nu_0=1$.

\begin{figure}[H]
\centering
\includegraphics[]{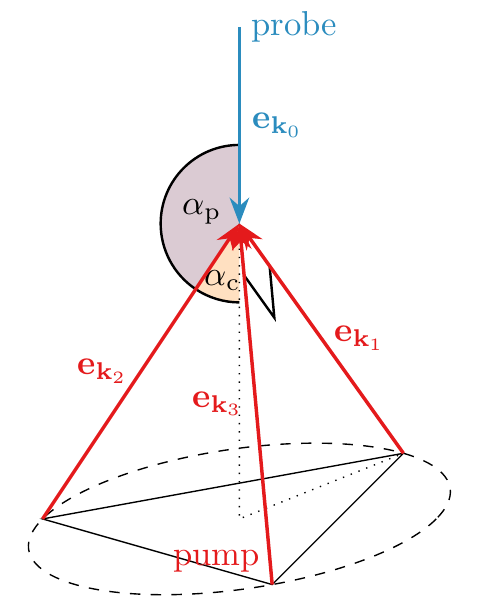}
\caption{Illustration of the setup. The three red arrows represent the unit wave vectors $\mathbf{e}_{\mathbf{k}_i}$ ($i\in\left\{1,2,3\right\}$) for the pump field. They form a right triangular pyramid where the isosceles are described by theses three unit wave vectors $\mathbf{e}_{\mathbf{k}_i}$. The angle between them are $90^{\circ}$ and the angle between these and the distance  perpendicular to the base is $\alpha_{\mathrm{c}}\approx54.74^{\circ}$. Besides, the blue arrow symbolizes the unit wave vector $\mathbf{e}_{\mathbf{k}_0}$ of the probe beam; it includes the angle $\alpha_{\mathrm{p}}\approx125.26^{\circ}$ with each pump unit wave vector.}
\label{fig:pyramid}
\end{figure}

Our aim is to generate a narrow high-intensity scattering center. By \newT{superimposing} laser \newT{fields} with different frequency and focusing them on the same spot coherently we try to construct such center. \newT{A} small scattering volume with intense field strength \newT{could be beneficial in achieving larger} values of scattering angles. \newT{Recently, it has been demonstrated that by} using the mechanism of coherent harmonic focusing (CHF) quantum vacuum signatures \newT{can be boosted substantially} \cite{Gordienko04,Gordienko05}. To make the signal photons distinguishable from the background photons of the driving laser beams we use a special three dimensional geometry to interfere the pump lasers. Former studies of CHF \newT{only} consider \newT{counter-propagating} laser beams along one axis \cite{Gordienko05,Karbstein19b}. \newT{Here, }we want to narrow down the volume of interaction by colliding pump lasers with different frequencies in a three dimensional geometry, see figure \ref{fig:pyramid}.       

For the pump laser beams we choose the wave vectors $\mathbf{k}_i=\nu_i \omega_0\, \mathbf{e}_{\mathbf{k}_i}$ with $i\in\left\{1,2,3\right\}$, where the unit wave vectors are
\begin{equation}
\mathbf{e}_{\mathbf{k}_1} = \left( - \sqrt{\frac{{2}}{{3}}},\, 0,\, \frac{1}{\sqrt{3}}\right)\,, \qquad \mathbf{e}_{\mathbf{k}_2} =  \left(\frac{1}{\sqrt{6}},\, - \frac{1}{\sqrt{2}},\, \frac{1}{\sqrt{3}}\right)\,, \qquad \text{and} \qquad  \mathbf{e}_{\mathbf{k}_3} = \left(\frac{1}{\sqrt{6}},\, \frac{1}{\sqrt{2}},\, \frac{1}{\sqrt{3}}\right)\,.
\end{equation}
\newT{The angle between two pump wave vectors is $90^{\circ}$, i.e., $\mathbf{e}_{\mathbf{k}_i}\cdot\mathbf{e}_{\mathbf{k}_j}=0$.} All pump \newT{beams are focused to the same spot which we } define as origin of the coordinate system. Furthermore, the angle between each beam and the $z$-axis is $\alpha_{\mathrm{c}}=\arctan\sqrt{2}$. The associated electric and magnetic fields point into the $\mathbf{e}_{\boldsymbol{\mathcal{E}_i}}$ and $\mathbf{e}_{\mathbf{B}_i}$ directions. The overall profile of each field \newT{amplitude} is given by the functions $\mathcal{E}_i\left(x\right)$. In our coordinate system, the field vectors for the $i$th pump beam are $\boldsymbol{\mathcal{E}}_i = \mathcal{E}_i\left(x\right) \mathbf{e}_{\boldsymbol{\mathcal{E}_i}}$ and $\mathbf{B}_i = \mathcal{E}_i\left(x\right) \mathbf{e}_{\mathbf{B}_i}$. We choose
\begin{equation}
\mathbf{e}_{\boldsymbol{\mathcal{E}}_1} = \left(\frac{1}{\sqrt{3}},\,0,\, \frac{\sqrt{2}}{\sqrt{3}}\right) \qquad \text{and} \qquad \mathbf{e}_{\boldsymbol{\mathcal{E}}_2} = \mathbf{e}_{\boldsymbol{\mathcal{E}}_3} = \left(\frac{\sqrt{2}}{\sqrt{3}},\,0,\, - \frac{1}{\sqrt{3}}\right)\,. 
\end{equation}
The unit vectors for the magnetic field are determined by $\mathbf{e}_{\mathbf{B}_i} = \mathbf{e}_{\mathbf{k}_i} \times \mathbf{e}_{\boldsymbol{\mathcal{E}}_i}$.

Now we want to probe the high-intensity region with the probe beam \newT{of} frequency $\omega_0$ and pulse duration $\tau$. To increase the signature of \newT{quantum vacuum} nonlinearity we want to maximize the angle between the probe beam and all pump beams. For the proposed setup the only option is to achieve that maximum angle by using the probe laser pointing towards the tip of the pyramid formed by the pump beams, see figure \ref{fig:pyramid}. We denote the wave vector of the probe field with $\mathbf{e}_{\mathbf{k}_0}=-\mathbf{e}_z$, it includes an angle $\alpha_{\mathrm{p}}$ with each pump field wave vectors $\mathbf{e}_{\mathbf{k}_i}$, $i\in\left\{1,2,3\right\}$. That angle is connected to $\alpha_{\mathrm{c}}$ by $\alpha_{\mathrm{p}}=\pi-\alpha_{\mathrm{c}}\approx 125.26^{\circ}$. In addition, we choose the polarization of the \newT{linear polarized} probe beam as $\mathbf{e}_{\boldsymbol{\mathcal{E}}_0} =  \mathbf{e}_y$.

We assume an alignment of all laser beams such that the maxima of intensity of each beam -- even the probe beam -- meet at the same point in spacetime. We define the collision center as the origin in our coordinate system. Each laser beam has a Gaussian profile. To boost the signal we focus all beams -- including the higher harmonics after frequency doubling -- to the same beam waist size $w_{i}=\lambda$ at the collision center.

\section{Results} \label{sec:results}
In this section we analyze the setup introduced in the previous section, calculate the differential number of signal photons analytically and discuss the advantages.     
\subsection{Derivation of the signal} \label{sec:DerivSig}
Let us compute the differential number of signal photons \newT{per shot} $\mathrm{d}^3N$ analytically. The signal amplitude $S_{\left(p\right)}\left(\mathbf{k}\right)$, see Eq. \eqref{eq:S1_FG}, yields
\begin{equation}
S_{\left(p\right)} \left(\mathbf{k}\right) = \frac{1}{\mathrm{i}}  \frac{e^2}{4\pi}  \frac{m_e^2}{45} \sqrt{\frac{k^{0}}{2}} \left(\frac{e}{m_e^2}\right)^3  \sum_{i,j,l=0}^3 \mathcal{I}_{ijl} \left(\mathbf{k}\right) g_{\left(p\right);ijl} \left(\hat{\mathbf{k}}\right) \label{eq:S_p}
\end{equation} 
with the Fourier integral
\begin{equation}
 \mathcal{I}_{ijl}\left(\mathbf{k}\right) \equiv \int\!\mathrm{d}^4x \mathrm{e}^{\mathrm{i}k_{\mu}x^{\mu}} \mathcal{E}_i\left(x\right)\mathcal{E}_j\left(x\right)\mathcal{E}_l\left(x\right)\,,  \label{eq:FourierInt}
\end{equation}
and an additional function $g_{\left(p\right);ijl}\left(\vartheta,\varphi\right)$ depending only on the signal photon angles $\vartheta$ and $\varphi$ and the polarization. This function is determined by the geometry of the unit vectors of all electromagnetic fields including the unit field vectors of the signal photon; we obtain 
\begin{equation}
g_{\left(1\right);ijl}\left(\vartheta,\varphi\right) = 2 \left( \mathbf{e}_{\left(1\right)} \cdot \mathbf{e}_{\boldsymbol{\mathcal{E}}_l}  - \mathbf{e}_{\left(2\right)} \cdot \mathbf{e}_{\mathbf{B}_l} \right) \left( \mathbf{e}_{\mathbf{B}_i} \cdot \mathbf{e}_{\mathbf{B}_j}  - \mathbf{e}_{\boldsymbol{\mathcal{E}}_i} \cdot \mathbf{e}_{\boldsymbol{\mathcal{E}}_j} \right) - \frac{7}{2} \left( \mathbf{e}_{\left(1\right)} \cdot \mathbf{e}_{\mathbf{B}_l}  + \mathbf{e}_{\left(2\right)} \cdot \mathbf{e}_{\boldsymbol{\mathcal{E}}_l} \right)  \left(\mathbf{e}_{\mathbf{B}_i} \cdot \mathbf{e}_{\boldsymbol{\mathcal{E}}_j} + \mathbf{e}_{\mathbf{B}_j} \cdot \mathbf{e}_{\boldsymbol{\mathcal{E}}_i} \right)\,, \label{eq:geo_func}
\end{equation}
and analogously $g_{\left(2\right);ijl}\left(\vartheta,\varphi\right)=\left.g_{\left(1\right);ijl}\left(\vartheta,\varphi\right)\right|_{\substack{\mathbf{e}_{\left(1\right)}\rightarrow\mathbf{e}_{\left(2\right)}\\\mathbf{e}_{\left(2\right)}\rightarrow-\mathbf{e}_{\left(1\right)}}}$.

The indices $i$,$j$,$l$ in the Fourier integral $\mathcal{I}_{ijl}\left(\mathbf{k}\right)$ and the geometry function $g_{\left(p\right);ijl}\left(\vartheta,\varphi\right)$ \newT{parameterize} all possible \newT{couplings of the driving laser field amplitudes} appearing in the signal photon amplitude. \newT{As the leading term to $\mathcal{L}_{\text{HE}}$ is quartic in the electromagnetic field, each signal photon $\gamma_{\left(p\right)}$ arises from the effective} \newTT{interaction} \newT{ of three laser fields: cf. Sec. \ref{sec:Theory} above.}

As mentioned in section \ref{sec:geo}, \newT{in order to model the amplitude profile $\mathcal{E}_i\left(x\right)$} we use a Gaussian beam profile in the limit of \newTT{infinite}  Rayleigh range \cite{Svelto10,Siegman86,Robertson1954}. Within this assumption, it can be represented as
\begin{equation}
\mathcal{E}_i \left(x\right) =  \frac{1}{2}A_i\, \mathcal{E}_{\star}\, \mathrm{e}^{-4 \frac{\left(r_i-t\right)^2}{\tau^2}}\, \mathrm{e}^{- \frac{x_{\perp,i}^2}{w_{i}^2\left(r_i\right)}}  \left( \mathrm{e}^{\mathrm{i} \nu_i\, \omega_0 \left(r_i -t \right)} + \mathrm{e}^{-\mathrm{i} \nu_i\, \omega_0 \left(r_i -t \right)}  \right) \,, \label{eq:E_x_infRay}
\end{equation} 
where we use the abbreviations $r_i = \mathbf{e}_{\mathbf{k}_i}\cdot\mathbf{x}$ and $x_{\perp,i}^2 = \left|\mathbf{e}_{\mathbf{k}_i}\times \mathbf{x} \right|^2$. The \newTT{infinite} Rayleigh range approximation is valid for weakly focused laser beams. This is \newT{particularly} well justified for pump laser beams generated by higher harmonics. 

Aiming \newT{at} observables, we use the signal amplitude $S_{\left(p\right)}\left(\mathbf{k}\right)$, see Eq. \eqref{eq:S_p}, together with the beam profile $\mathcal{E}_i\left(x\right)$ and the geometry introduced in section \ref{sec:geo} to calculate the differential number of signal photons
\begin{equation}
\mathrm{d}^3N_{\left(p\right)} \left(\mathbf{k}\right) = \mathrm{d} \mathrm{k} \mathrm{d} \cos\vartheta\, \mathrm{d}\varphi \frac{\mathrm{k}^2}{\left(2\pi\right)^3} \left|S_{\left(p\right)}\left(\mathbf{k}\right)\right|^2\,. \label{eq:d3N_result}
\end{equation}  
\newT{Moreover,} we can define a number density for photons in a given frequency range in between $\mathrm{k}_i$ and $\mathrm{k}_f$. This number density $\rho_{\left(p\right)}\left(\mathrm{k}_i,\mathrm{k}_f,\vartheta,\varphi\right)$ \newT{is obtained} after integration \newT{of Eq. \eqref{eq:d3N_result}} over this frequency range \newT{taking into account} the volume element $\mathrm{k}^2$:  
\begin{equation}
\rho_{\left(p\right)}\left(\mathrm{k}_i,\mathrm{k}_f,\vartheta,\varphi\right) = \frac{1}{\left(2\pi\right)^3} \int_{\mathrm{k}_i}^{\mathrm{k}_f} \!\mathrm{dk} \left|\mathrm{k}\, S_{\left(p\right)}\left(\mathbf{k}\right)\right|^2\,. \label{eq:rho_result}
\end{equation}
\newT{For an energy insensitive measurement of the signal photons we thus have} $\rho_{\left(p\right)}\left(\vartheta,\varphi\right)\equiv\rho_{\left(p\right)}\left(0,\infty,\vartheta,\varphi\right)$. Finally, we sum over both polarizations and integrate over the solid angles. This leads us to the total number of signal photons \newT{per shot}
\begin{equation}
N_{\text{tot}} = \sum_{p=1}^2 \int_0^{\infty}\!\mathrm{d}\varphi \int_{-1}^{1} \mathrm{d}\!\cos\vartheta \; \rho_{\left(p\right)} \left(\vartheta,\varphi\right) \,. \label{eq:N_tot}
\end{equation}

\subsection{Semi-analytic results} \label{sec:results_sa}
In the next step we want to use the above-mentioned formulae Eq. \eqref{eq:d3N_result} and Eq. \eqref{eq:rho_result} to derive results which can be measured in an actual experiment. The main focus lies on the distinguishability of the predicted signal photons from the background photons of the driving laser beams. First we provide estimates for the differential numbers of driving laser photons. Afterwards, we present the attainable numbers of signal photons encoding the signature of quantum vacuum nonlinearity based on the results derived in section \ref{sec:DerivSig}. 
\subsubsection{Driving laser beams} \label{sec:driving_laser}
In section \ref{sec:geo}, we have introduced a specific laser beam configuration allowing to create a narrow spatially confined scattering center of high intensity. This configuration is based on petawatt class lasers reaching strong electromagnetic field strengths. As we assumed Gaussian beam profiles, the far-field angular decay of the differential number of laser photons \newT{per shot} constituting a given driving laser beam follows as a Gaussian distribution. For the $i$th laser this quantity is given by \cite{Svelto10,Siegman86,Robertson1954}
\begin{equation}
\mathrm{d}^2 N_{i} = \mathrm{d}\varphi\,\mathrm{d}\cos\vartheta\;  \nu_i A_i^2 N_{\star} \mathrm{e}^{-2\nu_i^2\pi^2\vartheta_i^2\left(\vartheta,\varphi\right)}\,.
\end{equation}
Here, $\vartheta_i\left(\vartheta,\varphi\right)$ parameterizes the angular decay of the laser photons with respect to the unit wave vector $\mathbf{e}_{\mathbf{k}_i}$. The factor $N_{\star}=2\pi W/\omega_0$ is determined by the laser properties.

\subsubsection{Signal Photons}

To obtain the total number of signal photons \newT{per shot} $N_{\text{tot}}$, we have to combine the results for both polarizations; see Eq. \eqref{eq:N_tot}. Furthermore, we use the parameters encoding geometric and laser properties introduced in sections \ref{sec:geo} and \ref{sec:DerivSig} to determine the analytical expressions of $\mathrm{d}^3N_{\left(1,2\right)}$ and $\rho_{\left(1,2\right)}\left(\mathrm{k}_i,\mathrm{k}_f,\vartheta,\varphi\right)$. Using $\rho\left(\vartheta,\varphi\right)=\sum_{p=1}^2\rho_{\left(p\right)}\left(\vartheta,\varphi\right)$ we perform the integral over the solid angle numerically, which yields the total number of signal photons in the all-optical regime. We find $N_{\text{tot}}=325.29$ signal photons \newT{per shot} for the considered setup.

For an enhanced analysis we subdivide the frequencies of the resulting signal photons into several intervals, allowing for a spectrally resolved analysis of the signal. To this end, we use a frequency range $\mathrm{k}_i$ to $\mathrm{k}_f$ in the number density and integrate over the solid angles. We are in particular interested in the number of signal photons emitted in the frequency ranges of the driving laser beams. In table \ref{tab:ki_kf_N} we summarize the total numbers of signal photons \newT{per shot} associated with different frequency ranges.
\begin{table}[H]
\caption{Total number of signal photons \newT{per shot} attainable with the suggested setup based on three pump laser beams of frequencies $\omega_0=1.55\,\mathrm{eV}$, \newT{$2\omega_0=3.1\,\mathrm{eV}$ and $4\omega_0=6.2\,\mathrm{eV}$} and one probe beam of frequency $\omega_0=1,55\,\mathrm{eV}$. All beams are pulsed and feature a pulse duration of $\tau=25\,\mathrm{fs}$. Moreover, they are focused to a beam waist of $w_{i}=\lambda=800\,\mathrm{nm}$. We assume a one petawatt and a two petawatt laser at our disposal: one generates the pump fields (two petawatt) and one the probe (one petawatt). This table provides the number of signal photons for different frequency ranges $\mathrm{k}_i$ to $\mathrm{k}_f$.} \label{tab:ki_kf_N}
\centering
\begin{tabular}{ccc}
\toprule
\textbf{initial frequency }$\mathrm{k}_i$ \textbf{in} $\mathrm{eV}$	& \textbf{final frequency }$\mathrm{k_f}$ \textbf{in} $\mathrm{eV}$ 	& \textbf{number of signal photons} $N_{\text{tot}}$\\
\midrule
0.97			& 2.13			& 192.69\\
2.52			& 3.68			& 81.23\\
5.62			& 6.78			& 51.27\\
0.00			& $\infty$		& 325.29\\
\bottomrule
\end{tabular}
\end{table}

Moreover, we study the angularly resolved signal photon emission characteristics. A Mollweide projection allows us to transform the spherical data onto a flat chart. Because Mollweide projections do not change the areas of objects they are particularly suited to illustrate the spatial distribution of the signal photons. Note however, that these projections are not \newT{conformal} and thus do not conserve angles.
 
\begin{figure}[H]
\raggedright
\begin{minipage}{0.7\textwidth}
\raggedright
\includegraphics[scale=0.85]{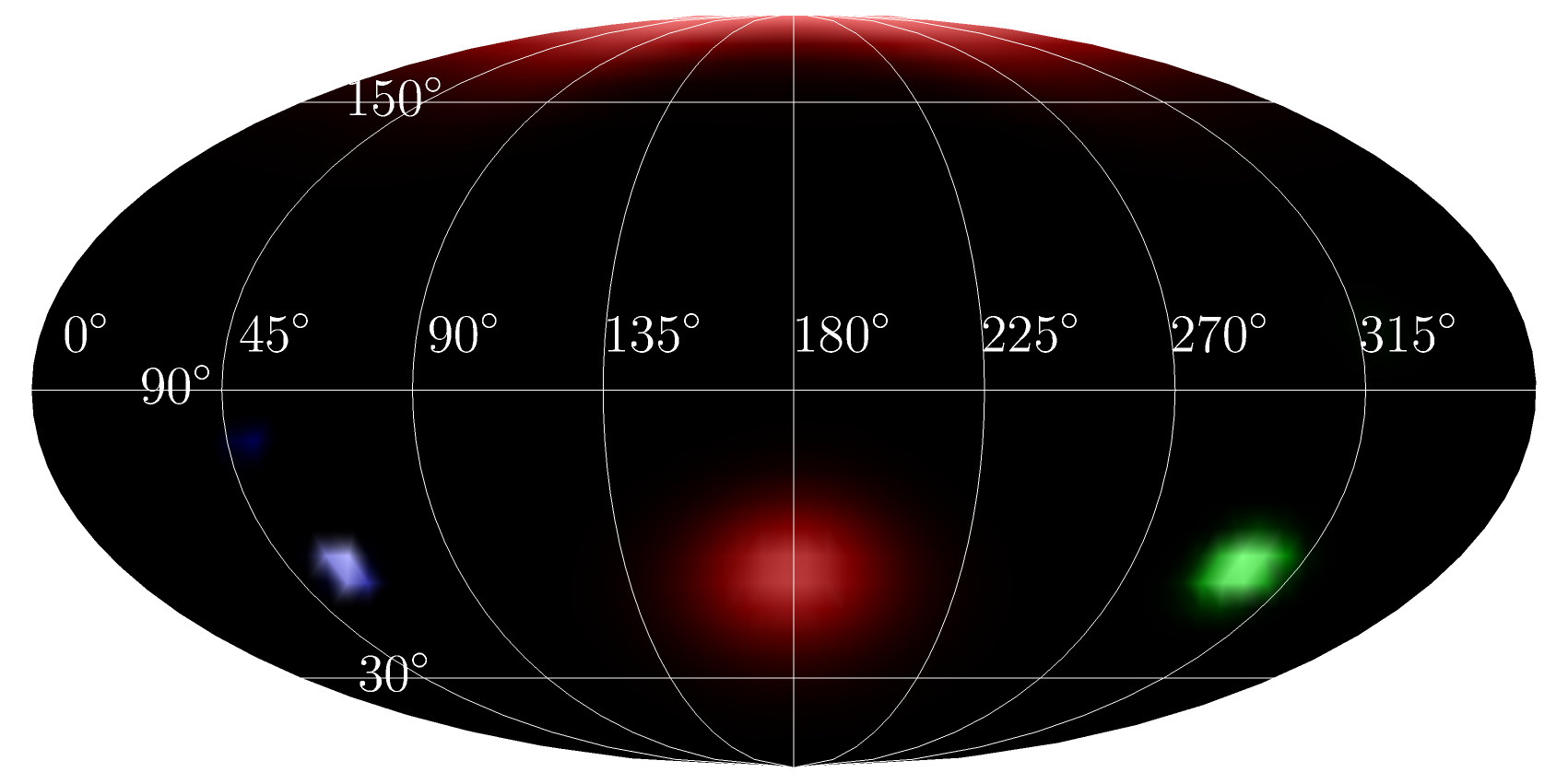}
\end{minipage}
\begin{minipage}{0.2\textwidth}
\raggedright
\includegraphics[scale=0.6]{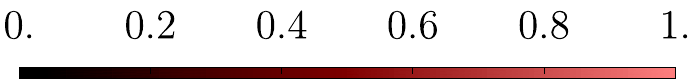}
\includegraphics[scale=0.6]{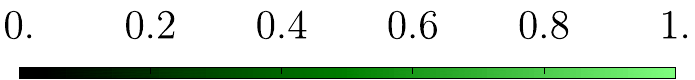}
\includegraphics[scale=0.6]{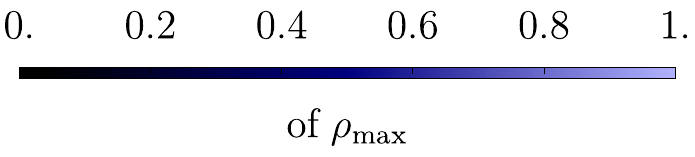}
\end{minipage}
\caption{Mollweide projection of the differential signal photon number $\rho\left(\mathrm{k}_i,\mathrm{k}_f,\vartheta,\varphi\right)$. The longitude gives the coordinate $\varphi$ and the latitude $\vartheta$. The three different colors denote the considered frequency regimes, i.e. $\mathrm{k}_{i,1}=0.97\,\mathrm{eV}$ to $\mathrm{k}_{f,1}=2.13\,\mathrm{eV}$ (red), $\mathrm{k}_{i,2}=2.52\,\mathrm{eV}$ to $\mathrm{k}_{f,3}=3.68\,\mathrm{eV}$ (green) and $\mathrm{k}_{i,3}=5.62\,\mathrm{eV}$ to $\mathrm{k}_{f,3}=6.78\,\mathrm{eV}$ (blue). The color scale is linear and normalized to the maximum values $\rho_{\text{max}}$ of each frequency regime. Next to the main peaks coinciding with the propagation directions of the driving beams, there are additional, less pronounced peaks in other directions.}
\label{fig:Mollweide_Signal}
\end{figure}

We present results for the spatial distribution of the signal photons for three frequency regimes, namely $\mathrm{k}_{i,1}=0.97\,\mathrm{eV}$ to $\mathrm{k}_{f,1}=2.13\,\mathrm{eV}$, $\mathrm{k}_{i,2}=2.52\,\mathrm{eV}$ to $\mathrm{k}_{f,3}=3.68\,\mathrm{eV}$, and $\mathrm{k}_{i,3}=5.62\,\mathrm{eV}$ to $\mathrm{k}_{f,3}=6.78\,\mathrm{eV}$. For each regime we determine $\rho\left(\mathrm{k}_i,\mathrm{k}_f,\vartheta,\varphi\right)$. Figure \ref{fig:Mollweide_Signal} shows these number densities. Here, the colors distinguish between different frequency regimes and the brightness indicates the relative number density. As signal photons of different frequencies are emitted into complementary directions, they can be depicted in one plot.

\subsubsection{Signal-to-background separation}
In the previous sections, we have studied the far-field distributions of both the driving laser photons and the signal photons encoding the signature of quantum vacuum nonlinearities. If we naively compare their total numbers, the signature of QED vacuum \newT{nonlinearity} seems to be undetectable in an experiment. The driving laser \newT{pulses consist} of the order of $10^{20}$ photons; the signal is made up of $325$ photons \newT{per shot}. However, taking into account additional properties of the signal we find possibilities to distinguish the signal from the background of the driving laser photons.

\begin{figure}[H]
\raggedright
\begin{minipage}{0.7\textwidth}
\raggedright
\includegraphics[scale=0.85]{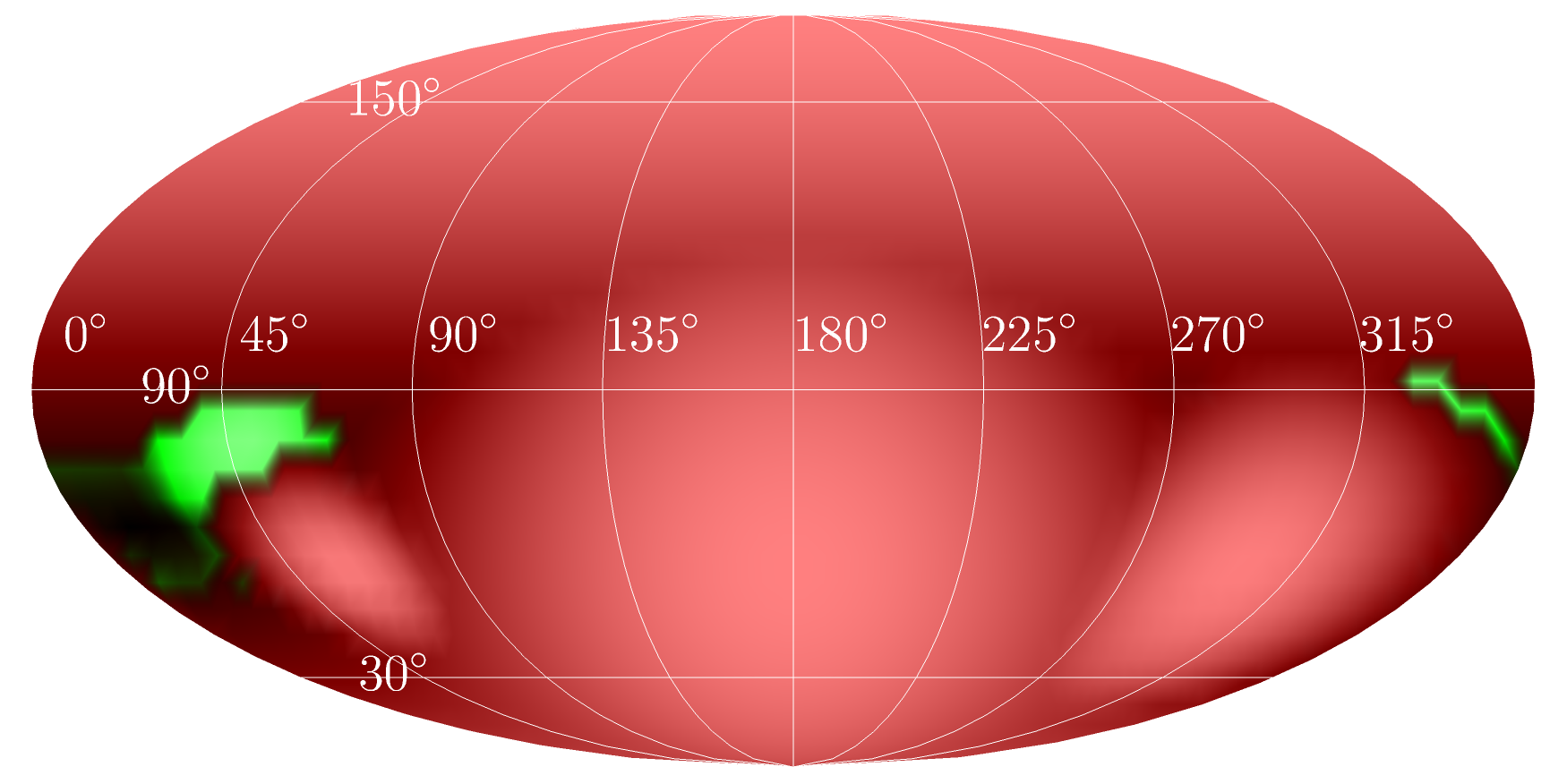}
\end{minipage}
\begin{minipage}{0.2\textwidth}
\raggedright
\includegraphics[scale=0.6]{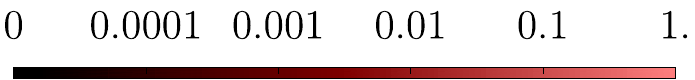}
\includegraphics[scale=0.6]{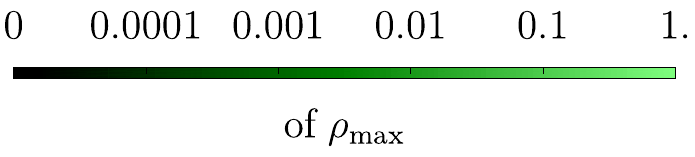}
\end{minipage}
\caption{Mollweide projection of the differential number of signal photons and driving laser photons in the all-optical regime. The longitude gives the coordinate $\varphi$ and the latitude $\vartheta$. In the red shaded areas the driving laser photons dominate, while in the green shaded areas the signal photons dominate. The color scale is logarithmic  and normalized to the maximum values $\rho_{\text{max}}$ of each type of signal.}
\label{fig:Mollweide_DiffLog}
\end{figure}

One possibility is the analysis of the spatial distribution of the photons \newT{constituting} the driving laser \newT{pulses} and the signal photons \newT{per shot}. The Mollweide projection in figure \ref{fig:Mollweide_DiffLog} highlights where the signal dominates over the driving laser photons. The driving laser photons dominate in the red shaded areas, while the signal dominates in the green shaded areas. Hence, in all green colored regions of figure \ref{fig:Mollweide_DiffLog} it is in principle possible to distinguish the signal photons from the background. In all frequency ranges, the main peaks in the signal photon distribution coincide with the directions of the driving laser beams. Besides, the signal photon distribution exhibits additional peaks. These peaks can be attributed to effective photon-photon interactions. With the suggested setup we manage to scatter signal photons into areas of lower driving laser intensity, i.e. areas with a much lower background. Using figure \ref{fig:Mollweide_Signal} we identify the frequency regime of the detectable signal photons. Our analysis implies that especially for the scattered signal photons of frequencies around \newT{$4\omega_0=6.2\,\mathrm{eV}$} the differential signal photon number surpasses the background. \newT{Correspondingly, focusing, e.g., on the far-field solid angle regime delimited by $\vartheta\in\left[80^{\circ},88^{\circ}\right]$ and $\varphi\in\left[40^{\circ},52^{\circ}\right]$ the signal photos should dominate over the background. We count 3.26 signal photons per shot in this regime. With a repetition rate of one shot per minute this should result in 195.6 discernible signal photons per hour. Taking into account the energy distribution in figure \ref{fig:Mollweide_Signal} we know that in this region the energy of the detected photons will be of the order of $4\omega_0=6.2\,\mathrm{eV}$. Besides this region, figure \ref{fig:Mollweide_DiffLog} shows that there are further angular regimes where the signal dominates over the background. This implies that state-of-the-art petawatt lasers collided and superimposed in a suitable configuration can induce signatures of photon-photon scattering accessible under realistic experimental conditions.} 
\section{Conclusions and outlook}
We have used the theoretical basis of QED in strong fields to derive analytical expressions for the differential numbers of signal photons encoding the signatures of quantum vacuum nonlinearity \newT{in experiments}. To achieve a measurable result we have introduced a special configuration based on two optical state-of-the-art petawatt lasers with frequency $\omega_0=1.55\,\mathrm{eV}$, pulse duration $\tau=25\,\mathrm{fs}$, and field energies $W=25\,J$ \newT{and $W_{\text{pump}}=50\,J$}. The pump laser beam was split into three different beams, two of which are transformed to higher frequencies $2\omega_0$ and $4\omega_0$ by means of higher harmonic generation \newT{accounting for experimentally realistic losses}. Upon aligning these beams in a right triangular pyramid with an angle of $90^{\circ}$ between each unit wave vector they \newT{form} the pump field. The second laser acts as a probe beam and propagates against the tip of that pyramid. We have derived analytical expressions accounting for the experimental parameters and loss factors and obtained the differential number of signal photons \newT{per shot} and the number density. After numerical evaluation we have compared these results with the background of the driving laser beams. We could in particular identify angular regimes where the differential signal photon number dominates the background, thereby constituting a prospective signature of QED nonlinearity in experiments.

The results \newT{discussed in this article} represent the current state of the analysis. Further \newT{analyses of the} properties of the signal are under investigation and will be published in the foreseeable future. One example is the spectral differential number, containing additional information beside the spatial distribution. In the latter, a widening of the spectral signal can be observed. The spectral width of the signal photons surpasses the spectral width of the driving lasers. In addition, we can change the beam properties and geometries for prospective studies, e.g. we can account for different loss factors.
Another interesting modification is to use different pulse durations or beam widths in the focus for the beams with different frequencies. Both of these quantities sensitively influence the scattering behavior of the signal photons.     

\vspace{6pt} 


\funding{This research was funded by the Deutsche
Forschungsgemeinschaft (DFG) under grant number 416611371 within the Research Unit FOR 2783/1.}

\acknowledgments{I thank Holger Gies, Felix Karbstein, Christian Kohlf\"urst, and Elena A. Mosman for the discussion and collaboration. In addition, I thank for the team of the Dubna Summer School 2019 ``Quantum Field Theory at the Limits: from Strong Fields to Heavy Quarks" with special thanks to David Blaschke and Mikhail Ivanov.}


%

\reftitle{References}


\externalbibliography{yes}
\bibliography{bab}




\end{document}